\documentstyle[12pt,aasms4]{article}

\journalid{}{}
\articleid{}{}

\begin{document}
\title{\bf \Large \ion{H}{2} Shells Surrounding Wolf-Rayet stars in 
M31}

\author{M. A. Bransford, D. A. Thilker, R. A. M. Walterbos\altaffilmark{1}, 
and N. L. King\altaffilmark{1}}
\affil{Department of Astronomy, New Mexico State University, Las Cruces, 
NM 88003}

\altaffiltext{1}{Visiting Astronomer, Kitt Peak National Observatory, National
Optical Astronomy Observatories, which is operated by the Association
of Universities for Research in Astronomy (AURA) under cooperative
agreement with the National Science Foundation.}

\begin{abstract}

We present the results of an ongoing investigation to provide a
detailed view of the processes by which massive stars shape the
surrounding interstellar medium (ISM), from pc to kpc scales. In this
paper we have focused on studying the environments of Wolf-Rayet (WR)
stars in M31 to find evidence for WR wind-ISM interactions, through
imaging ionized hydrogen nebulae surrounding these stars.

We have conducted a systematic survey for \ion{H}{2} shells
surrounding 48 of the 49 known WR stars in M31. There are 17 WR stars
surrounded by single shells, or shell fragments, 7 stars surrounded by
concentric limb brightened shells, 20 stars where there is no clear
physical association of the star with nearby H$\alpha$ emission, and 4
stars which lack nearby H$\alpha$ emission. For the 17+7 shells above,
there are 12 which contain one or two massive stars (including a WR
star) and that are $\leq$40 pc in radius. These 12 shells may be
classical WR ejecta or wind-blown shells. Further, there may be excess
H$\alpha$ point source emission associated with one of the 12 WR stars
surrounded by putative ejecta or wind-blown shells. There is also
evidence for excess point source emission associated with 11 other WR
stars. The excess emission may arise from unresolved circumstellar
shells, or within the extended outer envelopes of the stars
themselves.

In a few cases we find clear morphological evidence for WR shells
interacting with each other. In several H$\alpha$ images we see WR
winds disrupting, or punching through, the walls of limb-brightened
\ion{H}{2} shells.

\end{abstract}


\section{INTRODUCTION}

Wolf-Rayet (WR) stars are thought to be evolved, massive stars which
have lost a significant fraction of their outer envelopes. They are
sometimes surrounded by ring nebulae produced via the interaction of
the central star with its stellar ejecta or surrounding ambient medium
(e.g. Miller \& Chu 1993). Studies of the chemistry, morphology and
kinematics of Galactic WR ring nebulae have resulted in three main
categories (e.g. Chu 1981): (a) radiatively excited \ion{H}{2} regions
(R-type); (b) stellar ejecta (E-type); (c) and wind-blown bubbles
(W-type). A WR star does not play a primary role in the formation or
shape of an R-type shell (generally $>$ 30-40 pc in radius), only in
its ionization. An E-type shell is produced when the WR stellar wind
sweeps up mass lost from the star at an earlier epoch. Ejecta shells
tend to be small ($\sim$5-10 pc in radius; Chu 1981; Marston 1997),
and are generally identified chemically by high nitrogen-to-oxygen
ratios (indicative of CNO processing). The W-type shells ($\sim$10-20
pc in radius) are produced when a strong WR stellar wind sweeps up
ambient interstellar medium. Some Galactic WR stars have been observed
to exist within \ion{H}{1} shells (Arnal 1992). Eventually WR stars
undergo a supernova explosion at the end of their lives. WR stars,
therefore, have a substantial impact on their surroundings during the
various phases of their evolution.

We present the results of an ongoing study focusing on the interaction
of massive stars with the ISM, in nearby galaxies, from single stars
to associations (see Thilker, Braun, \& Walterbos 1998). Together,
these investigations provide a detailed view of the processes by which
massive stars shape the surrounding ISM, from pc to kpc scales. The
main goal of this paper is to study the environments of WR stars to
find evidence for WR wind-ISM interactions in M31, through imaging
ionized hydrogen nebulae surrounding these stars.

We have conducted a systematic search for \ion{H}{2} nebulae near WR
stars in M31. Our results are based on new and previously published
H$\alpha$, H$\alpha$+[\ion{N}{2}], and [\ion{S}{2}] imaging data of
M31 obtained at Kitt Peak National Observatory. M31 is attractive for
this type of study due its proximity (nebulae are resolvable down to
radii of $\sim$7 pc) and the fact that it provides a uniform sample of
WR stars at a well known distance (690 kpc). Surveys for WR stars in
M31, while not complete, have resulted in a catalog of 48 WR stars
(Massey \& Johnson 1998). Another WR star has recently been discovered
by Galarza, Walterbos, \& Braun (1999).

We present H$\alpha$ and B images of the 48 of these 49 WR stars which
fell within the area we imaged, and compile a catalog of ionized
hydrogen nebular ``shells'' in their vicinity. Deciding when an
\ion{H}{2} shell is physically associated with a WR star (i.e. an E-
or W-type shell) is not easy, since most WR stars are located close to
other massive stars and near classical \ion{H}{2} regions. Given the
small size of Galactic E-shells we can barely resolve the larger
shells of this type, but can easily resolve W- and R-type
shells. Although we cannot distinguish W- and R-type shells based
solely on imaging data, we attempt to determine which \ion{H}{2}
shells may be physically associated with WR stars based on criteria
discussed below. Excess point source H$\alpha$ emission associated
with WR stars might indicate unresolved E- or W-shells, or emission
from a WR stellar wind. The Galactic WNL star HDE 313846 has a broad
H$\alpha$ emission line, with an equivalent width of 23 ${\rm \AA}$,
which has been interpreted as arising in an extended outer envelope
indicating a strong stellar wind (Crowther \& Bohannan 1997). For this
reason we also list the stars with unresolved H$\alpha$ excess.

The paper is organized as follows. In $\S$2 and 3 we present details
of the observations and data reduction. In $\S$4 we describe our
morphological classification of H$\alpha$ emission near WR stars,
present H$\alpha$ and B images, and provide the physical properties of
all detected \ion{H}{2} shells. Further, we attempt to determine which
shells are causally associated with a WR star. We discuss the
environments of WR stars, and evidence for WR wind-ISM interactions,
in $\S$5. We present our summary and conclusions in $\S$6.

\section{THE OBSERVATIONS}

The data relevant to this study include previously published and new
imaging data of M31, both obtained at the KPNO 0.9-m telescope. The
earlier data were originally presented by Walterbos \& Braun (1992;
hereafter WB92). The WB92 data consist of 19 fields, located in the NE
half of M31, imaged through 27 ${\rm \AA}$ H$\alpha$ and [\ion{S}{2}]
filters, 75 ${\rm \AA}$ off-band continuum and broad-band B. Each
field measured 6$\farcm$7 $\times$ 6$\farcm$7.  Further details may be
found in WB92.

The new data, covering 8 fields mostly on the SW side of M31, were
obtained during the nights of 4-10 November 1997. The fields were
imaged through a pair of 75 ${\rm \AA}$ H$\alpha$+[\ion{N}{2}] and
off-band, red-shifted H$\alpha$ filters having nearly identical
bandpass shape, in addition to a blue (B-band) filter. A 75 ${\rm
\AA}$ filter had to be used since narrower filters were not available
to cover the radial velocities on the SW side of M31. Three to 7
individual frames were obtained for each H$\alpha$+[\ion{N}{2}],
off-band, and B-band field.  Total co-added exposure times for each
H$\alpha$+[\ion{N}{2}] and off-band field ranged from 2400 to 5400 s,
and for the B-band fields from 180 to 690 s.

The images were obtained using a TI 2k $\times$ 2k CCD detector.
Individual pixels measure 24 $\mu$m, which corresponds to 0$\farcs$68
or 2.3 pc at the distance of M31 (690 kpc). The fields measure
23$\farcm$2 on a side or, ignoring projection effects due to
inclination, 4.6 $\times$ 4.6 kpc$^{\rm 2}$. The typical resolution of
the WB92 data set was 2$\arcsec$, and for the 1997 data set
$\sim$1-2$\arcsec$ (which corresponds to a linear resolution of
$\sim$4-7 pc). Future papers will provide the entire 1997 data set and
its reduction, although we outline some essential reduction details
here.

\section{DATA REDUCTION AND CALIBRATION OF 1997 DATA SET}

Initial reduction of the individual frames (bias subtraction and flat
fielding) was performed within IRAF$\footnote{IRAF (Image Reduction
and Analysis Facility) is provided by the National Optical Astronomy
Observatories (NOAO). NOAO is operated by the Association of
Universities for Research in Astronomy (AURA), Inc. under cooperative
agreement with the National Science Foundation}$.  A twilight
``master-flat'' was constructed by co-adding twilight flats from
several different nights, in each filter. The twilight master-flats
successfully removed sensitivity variations across the chip. This
could be seen by the uniformity of the dome flats after correcting
them with the master-flats. Residual, large-scale variations were of
the order 1$\%$ across the field. Registered H$\alpha$+[\ion{N}{2}],
off-band, and blue images were then combined and cosmic rays were
removed, using the IRAF routine IMCOMBINE. The continuum images were
scaled to foreground stars in the H$\alpha$+[\ion{N}{2}]+continuum
images, and then subtracted to produce line emission images. The
images presented in this paper are each displayed with a logarithmic
scaling, to emphasize faint emission and structure within the
H$\alpha$ emission.

The line images were flux calibrated using the Oke \& Gunn (1983)
standard star HD 19445. We corrected for the slight effect of
H$\alpha$ absorption in the standard star. The rms noise after
calibration and continuum subtraction is $\sim$2 $\times$ 10$^{\rm
-17}$ ergs s$^{\rm -1}$ cm$^{\rm -2}$ pix$^{\rm -1}$. Expressed in
emission measure (the integral of the electron density {\it squared}
along the line of sight) this corresponds to 22 pc cm$^{\rm -6}$,
assuming an electron temperature of 10,000 K (note that for ionized
gas at 10,000 K, 5.6 $\times$ 10$^{\rm -18}$ erg s $^{\rm -1}$
cm$^{\rm -2}$ arcsec$^{\rm -2}$ = 2.8 pc cm$^{\rm -6}$
(e.g. WB92)). The accuracy of the calibration was tested by comparing
the fluxes of several normal \ion{H}{2} regions in overlapping fields
between the 1997 data set and WB92. The fluxes in the 1997 data are
higher by $\sim$30$\%$.  Since the WB92 data did not contain emission
from the [\ion{N}{2}]$\lambda\lambda$6548,6584 lines, and the ratio of
[\ion{N}{2}]/H$\alpha$ intensities was assumed to be be 0.3 in WB92,
agreement between the data sets is satisfactory.

In order to easily locate the positions of WR stars within our images,
the absolute orientation of the fields was determined using an
astrometery routine within IDL. This program was interactively given
the (X,Y) coordinates of several HST guide stars within our
fields. The known ($\alpha$,$\delta$) coordinates of the guide stars
were then used to determine a plate scale solution, giving each frame
a positional accuracy of $\sim$1$\arcsec$ relative to the guide star
reference frame.

\section{IDENTIFICATION AND PHYSICAL OF PROPERTIES OF \ion{H}{2} SHELLS}

We have classified the morphology of H$\alpha$ emission near WR stars
into two groups. The first category (Group I shells) consists of
\ion{H}{2} shells, or shell fragments, in which the WR star is at or
near the center of a single shell (Ia), or concentric limb brightened
shells (Ib). We consider Group I shells as potentially being causally
associated with a WR star.  Note that the multiple shells presented
here are probably not physically related to the multiple shells
reported by Marston (1995), since the former are generally E-type
shells, which are barely resolved in our data. The second category
(Group II stars) consists of WR stars where there is no clear causal
relation with nearby H$\alpha$ emission (IIa), or a lack of nearby
H$\alpha$ emission (IIb). Group IIa consists of WR stars that are
either near amorphous, diffuse H$\alpha$ emission, or on the edge of
\ion{H}{2} regions, and therefore have no preferred location with
respect to an \ion{H}{2} shell.  There are 17 WR stars surrounded by
Group Ia shells, 7 stars surrounded by Group Ib shells, 20 Group IIa
stars, and 4 Group IIb stars.

H$\alpha$ fluxes were computed for Group I shells, but not for nebular
emission near Group II stars. Polygonal masks (shown in Figure 1, see
below) used to compute the fluxes were chosen such that they most
likely contain all emission from a given \ion{H}{2} shell. Regions for
Group Ia shells are generally circular or elliptical in shape. The
multiple polygons for Group Ib shells are concentric in nature. The
interior polygons defined for Group Ib shells are used to compute the
fluxes of the inner shells, and the outer polygons are used to compute
the fluxes of the inner+outer shells. Outer shell fluxes then are
obtained as the difference of these two measurements. We defined
irregularly shaped regions for shell fragments, such that the boundary
traced the H$\alpha$ emission associated with a fragment. Masks
containing shell fragments therefore tend to be arc shaped. Pixels
within the regions defined above were summed to produce the total
shell+background fluxes. The background was estimated from one or more
polygonal regions immediately surrounding an \ion{H}{2} shell (not
shown), and were generally 2-3 times larger in area than the regions
used to compute the shell+background fluxes. The background was then
subtracted to produce a total shell H$\alpha$ flux.  Uncertainties in
the flux were estimated by adding in quadrature the rms background
uncertainty, for the number of source pixels, and root-N photon noise.

The sizes of shells were estimated by measuring several (5-10) radial
distances from a WR star to the outer edge of its surrounding shell,
and averaging these values. For a given \ion{H}{2} shell, the
uncertainties in the measurements varied depending on factors such as
the completeness and/or ellipticity of the shell. Radii were rounded
to the nearest 5 pc, and have typical uncertainties of $\pm$5pc.

Figure 1a contains H$\alpha$+[\ion{N}{2}] and blue (B-band) images of Group
Ia shells from the 1997 data set. The regions used to compute the
\ion{H}{2} shell fluxes are indicated. The B images serve as finder
charts and illustrate the spatial distribution of surrounding luminous
stars. Figure 1b is the same as Figure 1a, but contains images of
Group Ia shells from WB92. The H$\alpha$ images lack contribution from
[\ion{N}{2}]$\lambda\lambda$6548,6584 emission in this case. Group Ib
shells are shown in Figure 1c from the 1997 data set and in Figure 1d
from WB92. We note that the Group Ia shell surrounding MS11 appears in
Figure 1c, due to its proximity with MS12, which is surrounded by a
Group Ib shell. The names of the WR star(s) appearing in the images
are written above the blue images. The central WR star name appears
first whenever there is more than one WR star per image. Tables 1 \& 2
contain a listing of the properties of the \ion{H}{2} shells presented
in Figure 1, including the spectral subtype of the central star and
its coordinates, radii of the shells, and integrated H$\alpha$
luminosities (uncorrected for possible extinction).

Images of Group IIa stars are shown in Figure 2a, from the 1997 data
set, and Figure 2b, from WB92. Due to the proximity of ob33wr3 and
ob33wr2, the latter is shown in Figure 1c, although we consider this
star to belong to Group IIa. Group IIb stars are shown in Figure 2c,
from the 1997 data set.

The H$\alpha$ luminosity of Group I \ion{H}{2} shells, presented in
Figure 1, ranges between 1 and 43 $\times$ 10$^{\rm 36}$ ergs s$^{\rm
-1}$. In Figure 3 we show the distribution of H$\alpha$ luminosities
as a function of the size of the nebulae, and plot the same
distribution for the nebular structures surrounding WR stars in M33
(data from Drissen, Shara, \& Moffat 1991). Further we indicate the
detection limit in our data, with regard to shell-like objects.  The
minimum detectable luminosity for shells of varied size was determined
empirically through the use of a simple code for modeling the
anticipated appearance of idealized bubbles in a noisy background.
Our program initially computes the projected surface brightness
distribution of a shell (at very high spatial resolution) then
transforms to an observable form by convolving with an appropriate
Gaussian PSF and regridding.  Finally, we scale the simulated image to
units of emission measure.  This result is added to a realistic noise
background matching the properties of our observations.  Inspection of
the shell image with noise added determines whether or not a source of
the specified size and luminosity would be detected in our survey.
Although rather conservative, this procedure does not reflect
non-detection associated with confusion due to neighboring emission
line sources.  We suspect that the ambiguity related to objects
bordering the WR star may often dominate the ``error budget'' for
non-detection of WR shells.  For this reason, the detection limit
indicated in Figure 3 should be interpreted as a best-case scenario,
appropriate when the WR star is relatively isolated but otherwise just
a lower limit.

Figure 3 reveals that there is a gap between the detection limit and
the H$\alpha$ emission from Group I shells. Why don't we detect
fainter shells? One possible explanation is that the larger shells are
R-type, with many massive stars responsible for their ionization.
Typically Galactic WR shells expand at 10-50 km s$^{\rm -1}$ (Marston
1995), therefore a shell would need $\geq$4 $\times$ 10$^{\rm 5}$
years to expand to a radius of 20 pc. If the WR phase lasts 10$^{\rm
4}$-10$^{\rm 5}$ years (Marston 1995) shells larger than 20 pc would
no longer contain a WR star. The problem with this explanation is that
these timescales hold true only if the shell were to begin expanding
from the surface of the star. Shells, however, are likely to form far
from the surface of the star. In the intervening space between the
star and the radius at which the shell is formed, the stellar wind (or
possible ejecta) may expand at a much higher velocity than the
observed expansion velocity of the resulting shell. Further, one must
consider mass loss from pre-WR phases. The winds of massive stars can
operate on timescales of several Myr (the typical lifetime of an O
star), affecting the ISM up to radii of tens of parsecs. As for the
gap at small shell radii, there may be unresolved circumstellar shells
($\leq$ 5-7 pc) surrounding WR stars in our data (see below). The
point source detection limit is log L$_{\rm H\alpha}$ $\sim$ 34.5,
well below the scale in Figure 3. We note that the upper limit of
possible shell emission surrounding Group IIb stars also falls below
this scale. The gap therefore remains somewhat surprising.

The implied Lyman continuum luminosity of Group I shells ranges from
0.8 to 31.8 $\times$ 10$^{\rm 48}$ photons s$^{\rm -1}$. Esteban et
al. (1993) have estimated that Galactic WR stars can supply 4 to 37
$\times$ 10$^{\rm 48}$ photons s$^{\rm -1}$, indicating that all Group
I shells can in principle be ionized by a single WR star. However,
this is not proof that the only source of ionization is the WR
star. Spectral classification of the brightest blue stars within the
projected boundaries of the nebulae would be necessary to convincingly
resolve this issue.

We now discuss which Group I shells may be physically associated with
a solitary WR star. Our criteria are based on (a) the number of nearby
massive stars (other than the WR star), and (b) a comparison of the
sizes of the Group I shells with the sizes of WR nebulae in the LMC.

In Figure 4 we plot the sizes of Group I shells versus the number of
nearby, potentially ionizing stars (including the WR star in all
cases). Stars were selected if they were both (a) within one shell
radius (and therefore within the boundary of the shell or shell
fragment), and (b) have B - V colors $\leq$ 0 (photometry is taken
from Magnier et al. (1992)). Only stars bluer than B0 (B - V = -0.3)
can supply enough ionizing photons to be consistent with the H$\alpha$
luminosities of Group I shells. Our choice of B - V $\leq$ 0 was
prompted by considering the following two factors. First, there are
substantial uncertainties in the photometry from the Magnier
catalog. Second, we have made no reddening corrections. Thus the color
information on stars in the field is, unfortunately, crude. Note in
Figure 4 that Group I shells with radii less than $\sim$40 pc
generally contain one or two ionizing stars.

Rosado (1986) reported the radii of shells associated with (a) single
WR stars, (b) single OB stars, and (c) multiple OB \& WR stars in the
LMC. The mean radii are, respectively, 15$\pm$7 pc, 27$\pm$4 pc, and
69$\pm$33 pc (Rosado 1986). Single massive stars in the LMC are
surrounded by shells up to $\sim$30 pc in radius, consistent with the
radius estimated above for Group I shells

We suggest, allowing for chance alignment, those shells containing one
or two ionizing stars (including a WR star), or that are $\leq$ 40 pc
in radius, are possibly of E- or W-type. There are 5 Group Ia
shells that satisfy this criteria: ob69-F1, MS8, MS2, ob54wr1, and
ob42wr1. All 7 inner Group Ib shells satisfy this criteria, although
the outer shells for MS14 and MS4 do not. Therefore, 12 of the 48 WR
stars we observe (25$\%$) may be surrounded by ``classical'' WR E- or
W-type shells. The remaining 11 Group Ia shells contain three or more
stars, or have a radii $>$40 pc, and are probably R-type.

Previous observations have suggested Galactic E- and W-type shells are
generally $<$20 pc in radius (Chu 1981; Marston 1997). Our data (in
particular see Figure 4), on the other hand, suggest that the
\ion{H}{2} shells of this type {\it may} be larger in M31, on average,
than their Galactic counterparts. However, some of our larger shells
(including those classified as R-type) may be (and in some instances
probably are) classical, ring-like \ion{H}{2} regions or, for the very
largest shells, ``superbubbles'' (e.g. Oey \& Massey 1994; Oey 1996),
and may not be ``classical'' WR ring nebulae. Future observations to
specifically classify the \ion{H}{2} shells presented in this paper,
as either E-, W- or R-type, would be necessary to verify that E- or
W-shells are indeed larger in M31 than the Galaxy.

WR ring nebulae in the Galaxy and LMC have revealed interesting
statistical trends, suggesting WC stars generally are found within
nebulae of larger size than WN stars (Miller \& Chu 1993; Marston
1997; Dopita et al. 1994), possibly indicating an evolutionary
sequence. We compared the size distribution of the \ion{H}{2} shells
surrounding WN stars with those surrounding WC stars in M31. In light
of the discussion above, we restricted our comparison to shells
$\leq$40 pc in radius, and containing one or two massive stars
(i.e. those shells thought to be causally related to a single WR
star). For this particular subset of shells (3 contain WN stars and 14
contain WC stars) we found that shells containing a WN star were on
average smaller than those containing a WC star. If true, this
supports the evolutionary sequence, WN $\rightarrow$ WC (Marston
1997). However, given the small number of shells (in particular those
containing WN stars), our result is not statistically significant.
Further, considering all \ion{H}{2} shells in our sample, there are as
many WN stars within small shells (putative E- or W-type) as big
shells (putative R-type), as can be seen from Tables 1 \& 2. We
therefore caution that this result should be viewed as preliminary.

We conclude this section by discussing the excess point source
H$\alpha$ emission associated with a subset of WR stars in the 1997
and WB92 data. There are 12 stars with possible excess emission in the
H$\alpha$ passband. Tables 1 \& 2 list the luminosities and equivalent
widths of the excess emission for those WR stars within Group I
shells, and Table 3 for those stars in Group IIa. No stars within
Group IIb have excess emission. Excess H$\alpha$ emission associated
with WR stars might indicate unresolved E- or W-shells, or emission
from a WR stellar wind. We note that a small contribution from stellar
\ion{He}{2} $\lambda$6562 may also be present in the H$\alpha$ passband, 
but not enough to explain the equivalent widths of the excess
emission. The luminosities of the excess emission range from 3 to 10
$\times$ 10$^{\rm 34}$ ergs s$^{\rm -1}$, the equivalent widths from 5
to 240 ${\rm \AA}$. By way of comparison, the Galactic stellar ejecta
nebula RCW 58 has an H$\alpha$ luminosity of 1 $\times$ 10$^{\rm 35}$
ergs s$^{\rm -1}$ (Drissen et al. 1991), and WNL 313846 has a broad
H$\alpha$ emission line, with an equivalent width of 23 ${\rm \AA}$,
interpreted as arising in an extended outer envelop and indicative of
a strong stellar wind (Crowther \& Bohannan 1997). Crowther \& Smith
(1997) find that four LMC WN9-11 stars show nebular H$\alpha$ emission
lines (equivalent widths ranging from 32 to 83 ${\rm \AA}$)
superimposed on their stellar spectra, which is circumstellar and not
from an underlying \ion{H}{2} region.

\section{WR WIND-ISM INTERACTIONS}

Figure 1c contains an image of the Group Ia shell surrounding MS11 and
the Group Ib shell surrounding MS12. The bright star near the top of
the B-band image is a foreground star. MS11 is off-set from the center
of a large, limb-brightened bubble, and is part of a chain of OB
stars. To the south-west of MS11 is a multiple shell surrounding the
WR star MS12. MS12 is at the center of a small inner shell, and offset
to the west from the center of the outer shell. MS12 is the only
nearby ionizing star, which is supported by the flux of ionizing
photons implied by the H$\alpha$ luminosity of the surrounding shells.

Interestingly, there is morphological evidence that the shells
surrounding these stars may be interacting. Pointing radially away
from MS11 is a faint loop of H$\alpha$ emission. Perhaps not
coincidentally, the outer shell surrounding MS12 appears to be
impinging on the shell surrounding MS11, directly south of the
H$\alpha$ loop (or ``break-out'' feature), causing a noticeable
s-shaped distortion in the surface of the larger shell. The H$\alpha$
emission in the limb of the Group I shell is clearly enhanced in the
region to the north and south of the H$\alpha$ loop. An effect of the
expansion of these shells into one another may have been to weaken the
surface of the Group I shell, such that the fast wind from MS11 was
able to punch a hole in its wall. Interestingly, there is a gap
between the shells. Therefore, if the shells are interacting this
would imply they are radiation bounded, and the outer layers of the
shells are likely neutral.

In Figure 1d we see, for ob33wr3 and ob33wr2, another interesting case
of potentially interacting shells, and the effects of WR winds. A
large arc of emission south-west of ob33wr3 is impinging upon a shell
structure containing ob33wr2. Interestingly, the Group Ib shell is
centered within this arc. As above, the gap between the shell and this
arc could indicate the shell is radiation bounded, with neutral
material in between. The possibility of a neutral outer layer is
strengthened by the evidence for a dust lane, seen in the B image,
superimposed on the gap. The probable action of a fast wind arising
from ob33wr2 is evident from a small cavity (blown?) in the wall this
shell, south-east of the star. An arc is seen surrounding ob10wr1 (see
Figure 1d) perhaps being driven into ob10 by the expansion of this
Group Ib shell. Finally, the potential action of a WR wind can be seen
for the case of the Group IIa star MS17 (see Figure 2a), where it has
apparently disrupted the wall of a diamond shaped \ion{H}{2} shell, in
which it is embedded.

Obtaining detailed kinematics (in both the ionized and neutral gas) of
these interacting shells, and the gas effected by WR winds, may prove
extremely useful in understanding the dynamical evolution of shell
structures in the ISM of galaxies. Further, in addition to
spectroscopy, narrow-band imaging, in lines such as
[\ion{O}{3}]$\lambda$5007 or [\ion{O}{1}]$\lambda$6300, may prove
useful in detecting shock fronts in a wind-ISM interaction region.
Will shells merge to become larger single structures?  What role do WR
winds play in enlarging pre-existing shell structures?  Are the
shells, presented in this paper, the evolutionary precursors of
\ion{H}{1} shells (or supershells). What role do WR stars play in
producing a frothy \ion{H}{1} medium in galaxies?

\section{SUMMARY AND CONCLUSIONS}

We have presented the results of an ongoing investigation to provide a
detailed view of the processes by which massive stars shape the
surrounding ISM, from pc to kpc scales. In this paper we have focused
on studying the environments of WR stars to find evidence for WR
wind-ISM interactions in M31, through imaging ionized hydrogen nebulae
surrounding these stars.

We have classified the morphology of the H$\alpha$ emission near 48 WR
stars into two groups, Group I and Group II. The first category (Group
I shells) consists of \ion{H}{2} shells, or shell fragments, in which
the WR star is at or near the center of a single shell (Ia), or
concentric limb brightened shells (Ib). We consider Group I shells as
potentially being causally associated with a WR star (i.e. E- or
W-type shells). There are 17 WR stars within Group Ia shells, and 7
stars within Group Ib shells. The second category (Group II stars)
consists of WR stars where there is no clear causal relation with
nearby H$\alpha$ emission (IIa), or a lack of nearby H$\alpha$
emission (IIb). Group IIa consists of WR stars that are either near
amorphous, diffuse H$\alpha$ emission, or on the edge of \ion{H}{2}
regions, and therefore have no preferred location with respect to an
\ion{H}{2} shell. There are 20 Group IIa stars, and 4 Group IIb
stars. Of the 48 WR stars in our sample, 24 are surrounded by Group I
shells (50$\%$) and 24 are Group II stars (50$\%$).

We suspect that Group I shells which contain one or two massive stars,
or are $\leq$ 40 pc in radius, may be E- or W-type shells. There are 5
Group Ia shells that satisfy this criteria: ob69-F1, MS8, MS2,
ob54wr1, and ob42wr1. All 7 inner Group Ib shells satisfy these
criteria, although the outer shells for MS14 and MS4 do not.
Therefore, 12 of the 24 WR stars within Group I shells (or 25$\%$ of
the sample of 48) appear surrounded by resolved E- or W-type shells.

There are 12 WR stars (25$\%$ of the sample) that have excess emission
within the H$\alpha$ bandpass (one WR star with excess emission is in
common with the 12 WR stars surrounded by putative E- or W-type
shells). The excess H$\alpha$ emission may arise from unresolved E- or
W-type shells, or in an extended outer envelope indicating a strong
stellar wind. Excess unresolved emission occurs for stars within Group
I shells, and for Group IIa stars. No stars within Group IIb have
excess emission. If we assume that the unresolved excess H$\alpha$
emission arises from an unresolved E- or W-type shell, nearly one-half
of our sample may have surrounding E- or W-type shells.

We find morphological evidence that shells surrounding MS11 and MS12,
and ob33wr3 and ob33wr2, may be interacting. Interestingly, there are
gaps between the shells. Therefore, if these shells are interacting
this implies they are radiation bounded, and the outer layers of the
shells are likely neutral. An arc is seen surrounding ob10wr1 perhaps
being driven into ob10 by the expansion of this Group Ib
shell. Obtaining detailed kinematics (in both the ionized and neutral
gas) of these interacting shells, and the gas effected by WR winds,
may prove extremely useful in understanding the dynamical evolution of
shell structures in the ISM of galaxies.

{\bf Acknowledgements}

We would like to thank Phil Massey for a useful conversation when
visiting NMSU. This research was partially supported by the National
Science Foundation through grant AST 96-17014.

%
%

%
%

\clearpage

\begin{figure}
\figurenum{1a}
\figcaption[wr.fig1a.ps]{H$\alpha$+[\ion{N}{2}] and blue (B-band) images 
of Group Ia shells from the 1997 data set. The regions used to compute
the \ion{H}{2} shell fluxes are indicated. The B images serve as
finder charts and illustrate the spatial distribution of surrounding
luminous stars. The names of the WR star(s) appearing in the images
are written above the blue images. The central WR star name appears
first whenever there is more than one WR star per image. The images
measure 300 $\times$ 300 pc$^{\rm 2}$.}
\end{figure} 

\begin{figure}
\figurenum{1b} 
\figcaption[wr.fig1b1.ps,wr.fig1b2.ps]{Same as Figure 1a, but contains 
images of Group Ia shells from WB92. The H$\alpha$ images lack
contribution from [\ion{N}{2}]$\lambda\lambda$6548,6584 emission in
this case.}
\end{figure}

\begin{figure}
\figurenum{1c}
\figcaption[wr.fig1c.ps]{Same as Figure 1a, but shows Group Ib shells
from the 1997 data set.}
\end{figure}

\begin{figure}
\figurenum{1d}
\figcaption[wr.fig1d.ps]{Same as Figure 1b, but shows Group Ib shells
from WB92.}
\end{figure}

\begin{figure}
\figurenum{2a}
\figcaption[wr.fig2a.ps]{Images of Group IIa stars from 1997 data set.}
\end{figure}

\begin{figure}
\figurenum{2b}
\figcaption[wr.fig2b.ps]{Images of Group IIa stars from WB92.}
\end{figure}

\begin{figure}
\figurenum{2c}
\figcaption[wr.fig2c.ps]{Images of Group IIb stars from 1997 data set.}
\end{figure}

\begin{figure}
\figurenum{3} 
\figcaption[wr.fig3.ps]{Distribution of H$\alpha$ luminosities as a function 
of the size of the nebulae. Also shown is the same distribution for the 
nebular structures surrounding WR stars in M33 (data from Drissen, Shara, 
\& Moffat 1991). Further we indicate the detection limit in our data, with 
regard to shell-like objects.}
\end{figure} 

\begin{figure}
\figurenum{4} 
\figcaption[wr.fig4.ps]{Radii of Group I shells versus the 
number of nearby, potentially ionizing stars (including the WR star in all
cases).} 
\end{figure}

\end{document}